\documentclass[12pt]{article}
\usepackage{geometry}
\usepackage{makeidx}
\usepackage[T1]{fontenc}
\usepackage[dvipsnames,svgnames,table]{xcolor}
\usepackage{epstopdf}
\usepackage{epsfig}
\usepackage{textcomp}
\usepackage{hyperref}
\usepackage{amsmath}
\usepackage{amssymb}
\usepackage{multirow}
\usepackage{multicol}
\usepackage{booktabs}
\usepackage{lastpage}
\usepackage{hyperref} 
\usepackage{graphicx}
\usepackage{enumerate}
\usepackage{threeparttable}
\usepackage{float}
\usepackage{array}
\usepackage{cite}
\usepackage{color,soul}
\usepackage{listings}
\usepackage{tikz}
\usepackage{longtable}
\makeatletter
\renewcommand\section{\@startsection{section}{1}{\z@}%
   {-2.5ex \@plus -1ex \@minus -.2ex}%
   {2.3ex \@plus.2ex}%
   {\normalfont\large\bfseries}}
\renewcommand\subsection{\@startsection{subsection}{1}{\z@}%
   {-2.5ex \@plus -1ex \@minus -.2ex}%
   {2.3ex \@plus.2ex}%
   {\small\bfseries}}

\begin{document}
\bigskip
\title{\textbf{Effect of hydrogen on albedo neutrons in lunar surface: A computational investigation by means of GEANT4 simulations}}
\medskip
\author{\small A. Ilker Topuz$^{1}$}
\medskip
\date{\small$^1$Manipal Centre for Natural Sciences, Centre of Excellence, Manipal Academy of Higher Education, Manipal 576104, India\\aitopuz@protonmail.com}
\maketitle
\begin{abstract}
The remote sensing observations of the lunar soil by using neutron spectroscopy as well as infrared spectroscopy in various studies suggest the presence of hydrogen in the lunar surface, particularly around the lunar polar regions. Additionally, the cosmic ray protons incident on the lunar surface induce the generation of secondary neutrons depending on the composition of the surface soil and the associated density, a certain number of which, called albedo neutrons, leak from the lunar regolith. Motivated by the interaction of the cosmic ray protons with the lunar surface in the absence and presence of hydrogen, a series of GEANT4 simulations are employed in the current study to unveil the influence of hydrogen on the energy spectrum of the albedo neutrons that escape from the lunar surface. Initially, a discrete 69-bin proton energy spectrum between 0.4 and 115 GeV based on the PAMELA spectrometer is implemented into GEANT4. Subsequently, a single-volume lunar surface of 2-m thickness is constructed where it is assumed that a single layer of either 1.6 or 1.93 g/cm$^{3}$ constitutes the lunar regolith. The material composition in the present study includes 43.7 wt.$\%$ oxygen, 0.3 wt.$\%$ sodium, 5.6 wt.$\%$ magnesium, 9 wt.$\%$ aluminum, 21.1 wt.$\%$ silicon, 8.5 wt.$\%$ calcium, 1.5 wt.$\%$ titanium, 0.1 wt.$\%$ manganese, and 10.2 wt.$\%$ iron in accordance with another study. Then, 0.1 wt.$\%$ hydrogen is introduced by replacing oxygen in order to assess the impact of hydrogen on the secondary neutrons. In the wake of irradiating a lunar surface of 3$\times$2$\times$3 m$^{3}$ with a planar vertical PAMELA proton beam of 20$\times$20 cm$^{2}$,  the depth profile of the generated neutrons as well as the corresponding initial energy spectrum is primarily obtained in the absence and presence of 0.1 wt.$\%$ hydrogen by voxelizing the entire geometry with 100 cells of 2-cm thickness. Next, a surface detector is placed at the top of the lunar surface in order to collect the albedo neutrons in both cases, and the energy spectra of the albedo neutrons are acquired in terms of thermal, epithermal, and fast neutrons. From the GEANT4 simulations in this study, it is shown that the presence of 0.1 wt.$\%$ hydrogen is observable in each energy regime of the albedo neutrons at the lunar surface, thereby providing an indication about the elemental variation of the lunar soil. 
\end{abstract}
\textbf{\textit{Keywords: }} Cosmic ray protons; PAMELA spectrometer; Albedo neutrons; Lunar surface; GEANT4.
\section{Introduction}
\label{Intro}
The exploration of the lunar surface pertaining to the presence and distribution of hydrogen~\cite{watson1961possible,anand2010lunar,liu2012direct} is of particular importance for the future lunar exploration missions since the remote sensing techniques like neutron and infrared spectroscopy ~\cite{hoshino2020lunar, li2018direct, sanin2017hydrogen} already hint at the existence of hydrogen in the lunar polar regions~\cite{feldman2000polar,feldman2001evidence,cocks2002lunar,fisher2017evidence}. In this regard, the cosmic ray protons that bombard the lunar soil trigger the generation of secondary neutrons contingent on the composition of the lunar soil, and a specific portion of these secondary neutrons that are called albedo neutrons leak from the lunar regolith~\cite{zaman2022modeling}. Thus, the emission of the secondary neutrons from the lunar surface interacting with the cosmic ray protons provides valuable insights into the composition of the lunar regolith, and the albedo neutrons that escape from the lunar surface serve as a key indicator of surface composition.

In this study, the influence of hydrogen existing in the lunar soil on the energy spectrum of the albedo neutrons is investigated by means of GEANT4 simulations~\cite{agostinelli2003geant4, TopuzGithubBulkLunar}. To achieve this aim, a discrete 69-bin proton energy spectrum between 0.4 and 115 GeV based on the PAMELA spectrometer~\cite{adriani2011pamela} is used through a probability grid that is already implemented into GEANT4~\cite{topuz2022dome, topuz2023particle}. A 2-m-thick lunar surface of 3$\times$2$\times$3 m$^{3}$ is built by assuming a single bulky volume with a density of 1.6~\cite{naito2023global} or 1.93~\cite{li2017simulation} g/cm$^{3}$. The same material composition is defined for both these densities, and this material composition consisting of 43.7 wt.$\%$ oxygen, 0.3 wt.$\%$ sodium, 5.6 wt.$\%$ magnesium, 9 wt.$\%$ aluminum, 21.1 wt.$\%$ silicon, 8.5 wt.$\%$ calcium, 1.5 wt.$\%$ titanium, 0.1 wt.$\%$ manganese, and 10.2 wt.$\%$ iron is extracted from another study~\cite{zaman2022modeling}. In the cases where hydrogen is present, 0.1 wt.$\%$ hydrogen is added into the material composition by removing an equivalent amount of oxygen. By injecting a vertical planar proton beam of 20$\times$20 cm$^{2}$ founded on the PAMELA spectrometer, the generation depth as well as the initial energy spectrum of the generated neutrons within the voxelized lunar soil is first determined in the absence and presence of 0.1 wt.$\%$ hydrogen. Then, a surface detector is introduced in order to record the albedo neutrons leaking out the lunar surface without and with 0.1 wt.$\%$ hydrogen, and the acquired neutron energy spectra are divided into three energy ranges, namely thermal ($E\le1$ eV), epithermal (1 eV $< E\le1$ keV), and fast ($E > 1$ keV)~\cite{NASANeutronranges2024}. This study is organized as follows. In section~\ref{Protonenergyspectrum}, the energy spectrum of the incident cosmic ray protons delivered by the PAMELA spectrometer is exhibited, and the implementation of the 69-bin discrete energy spectrum into GEANT4 is briefly elucidated. In section~\ref{Simulation_setup}, the features of the lunar surface in the present study are stated in addition to the properties of the GEANT4 simulations, while the current simulation results are shown in conjunction with the neutron generation depth, the generated neutron spectrum, and the albedo neutron spectrum by excluding and including 0.1 wt.$\%$ hydrogen in section~\ref{Simulation_outcomes}. Finally, section~\ref{Conclusion} incorporates  the conclusions drawn from the present GEANT4 simulations.
\section{PAMELA proton energy spectrum}
\label{Protonenergyspectrum}
\begin{figure}[H]
\begin{center}
\includegraphics[width=8.25cm]{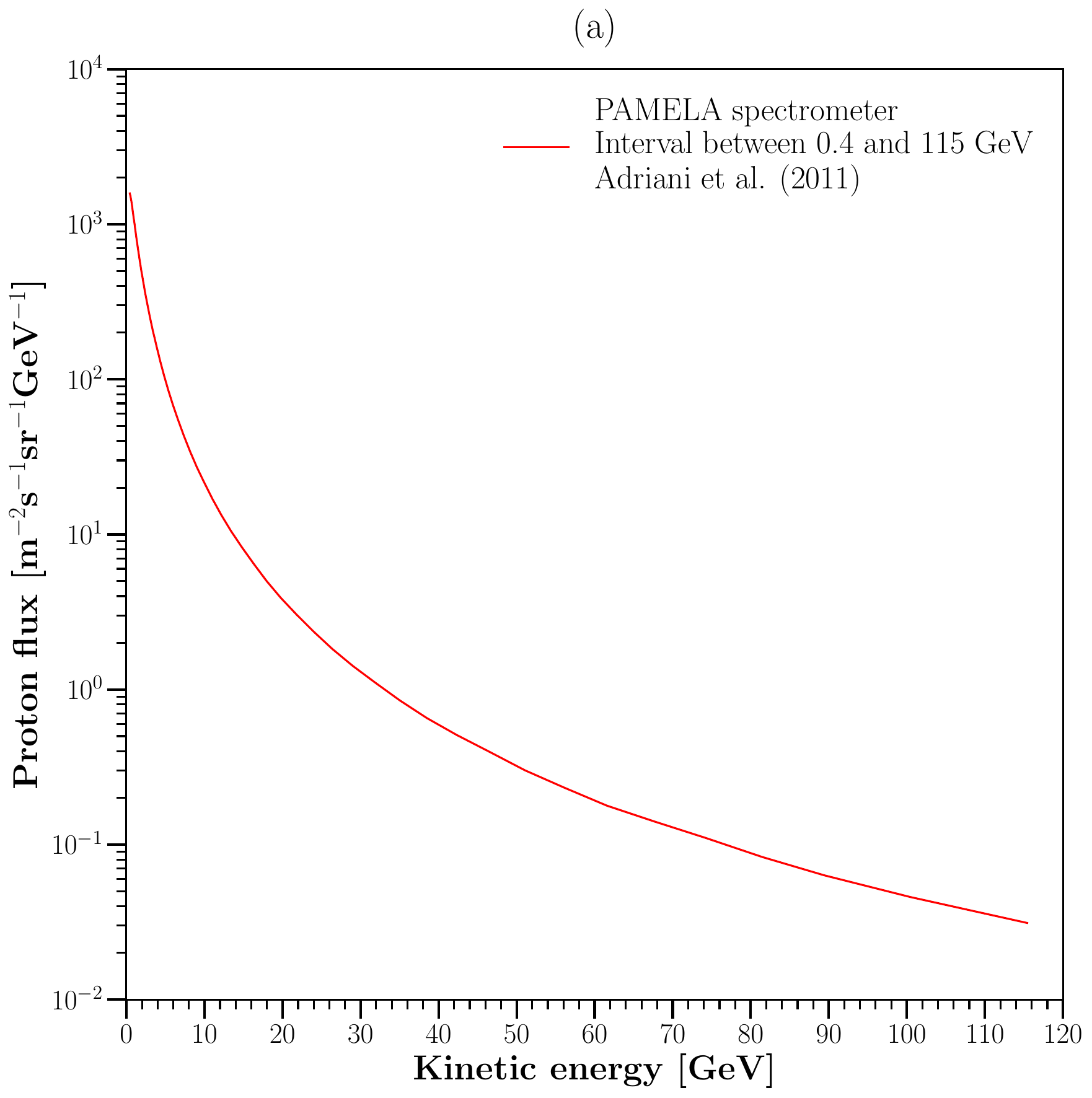}
\includegraphics[width=8.25cm]{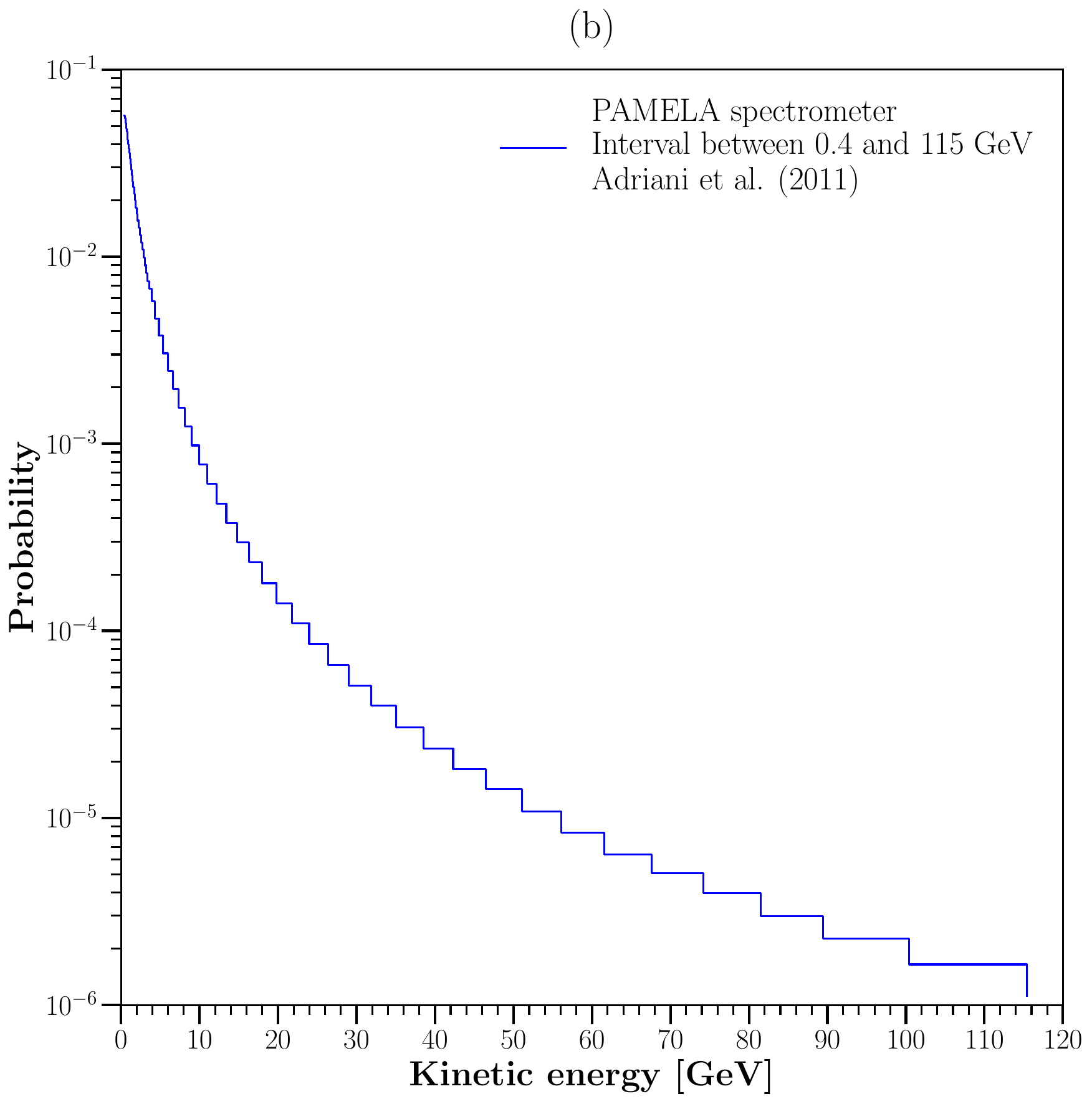}
\caption{PAMELA proton spectrum between 0.4 and 115 GeV (a) particle flux and (b) discrete probabilities.}
\label{PAMELA_proton_spectrum} 
\end{center}
\end{figure}

In the present study, the lunar surface is irradiated by bombarding it with cosmic ray protons to generate secondary neutrons via the interaction of the cosmic ray protons with the lunar soil. To accomplish this objective, among the existing experimental cosmic ray proton spectra is the PAMELA proton spectrum that provides the proton flux values at the discrete energy bins between 0.4 and 1009.4 GeV~\cite{adriani2011pamela}. According to the PAMELA proton spectrum, the cosmic ray proton flux drastically diminishes when the kinetic energy of the cosmic ray protons increases, leading to a proton flux ratio in the order of 10$^{-5}$ between 0.4 and 115 GeV as described in Fig.~\ref{PAMELA_proton_spectrum}(a). Since the contribution of the particle population after 115 GeV is negligible, a threshold value of 115 GeV is used in the current study.
The tabular proton flux values from the PAMELA spectrometer permit to calculate the discrete probabilities for each energy bin. In order to implement the PAMELA cosmic ray proton spectrum into GEANT4, the corresponding discrete probability of each energy bin is therefore computed by adding up all the energy bins from 0.4 to 115 GeV and dividing each energy bin by the total sum. Fig.~\ref{PAMELA_proton_spectrum}(b) illustrates the variation of the discrete probabilities determined from the PAMELA spectrometer with respect to the kinetic energies. For the purpose of implementing the discrete proton energy spectra obtained from the PAMELA spectrometer, a strategy to inject the incoming protons is subsequently integrated by means of G4ParticleGun as can be found in the previous studies~\cite{topuz2022dome, topuz2023particle}. By recalling the unity condition, a grid is built by summing up the discrete probabilities, the interval of which starts with 0 and ends in 1. Thus, each cell in this grid, i.e. the difference between two points on the probability grid, specifies a discrete probability. Then, a random number denoted by $\xi$ between 0 and 1 is generated by using the pre-defined uniform number generator called G4UniformRand(). Finally, this random number is scanned on the probability grid by checking the difference between the grid points, and the particular discrete energy is assigned when the random number matches with the associated cell. 
\section{Simulation setup}
\begin{figure}[H]
\begin{center}
\includegraphics[width=14cm]{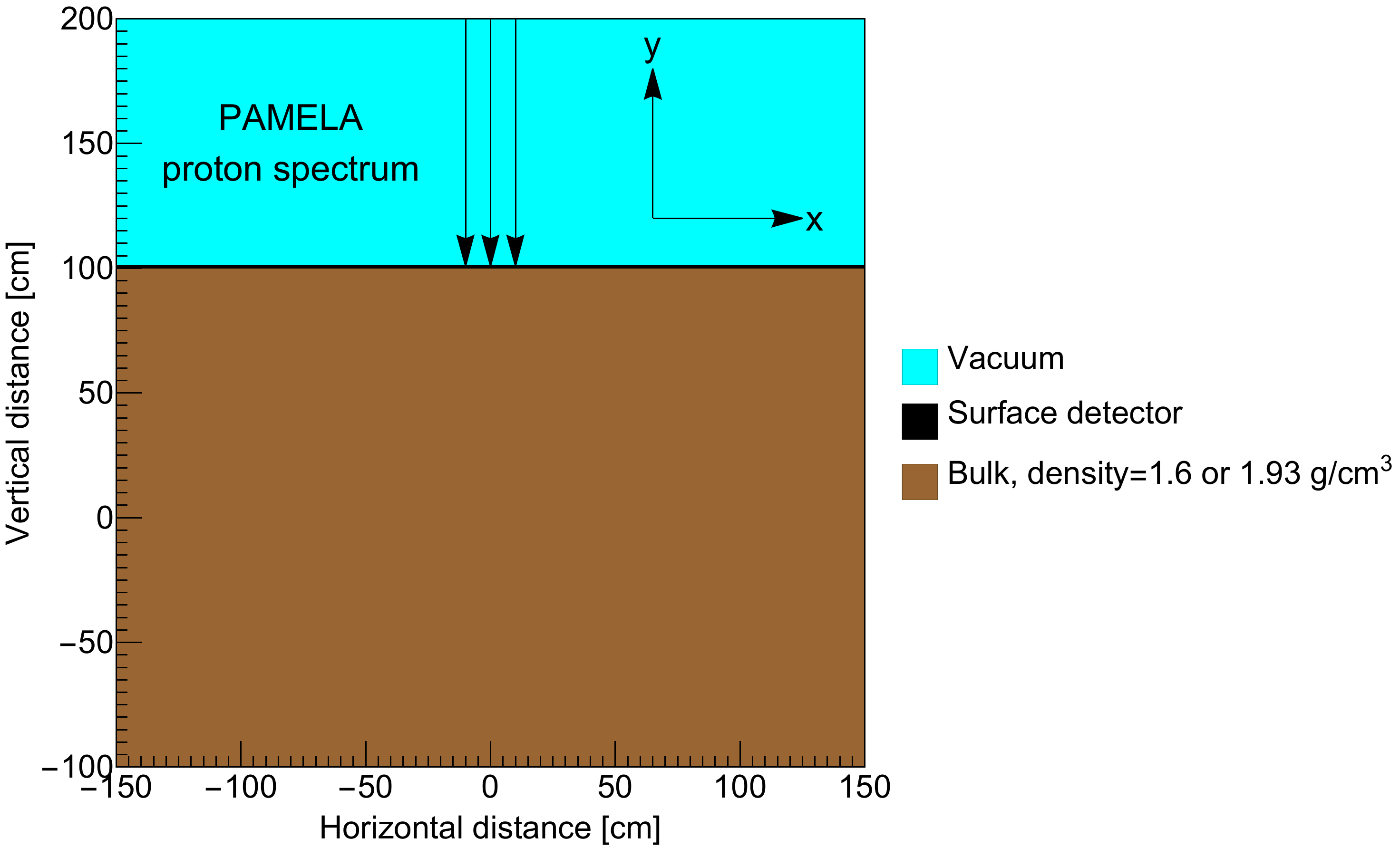}
\caption{Geometrical and structural layout of 2-m-thick lunar surface in GEANT4 simulations.}
\label{Lunar_surface_model}
\end{center}
\end{figure}

\begin{figure}[H]
\begin{center}
\includegraphics[width=11cm]{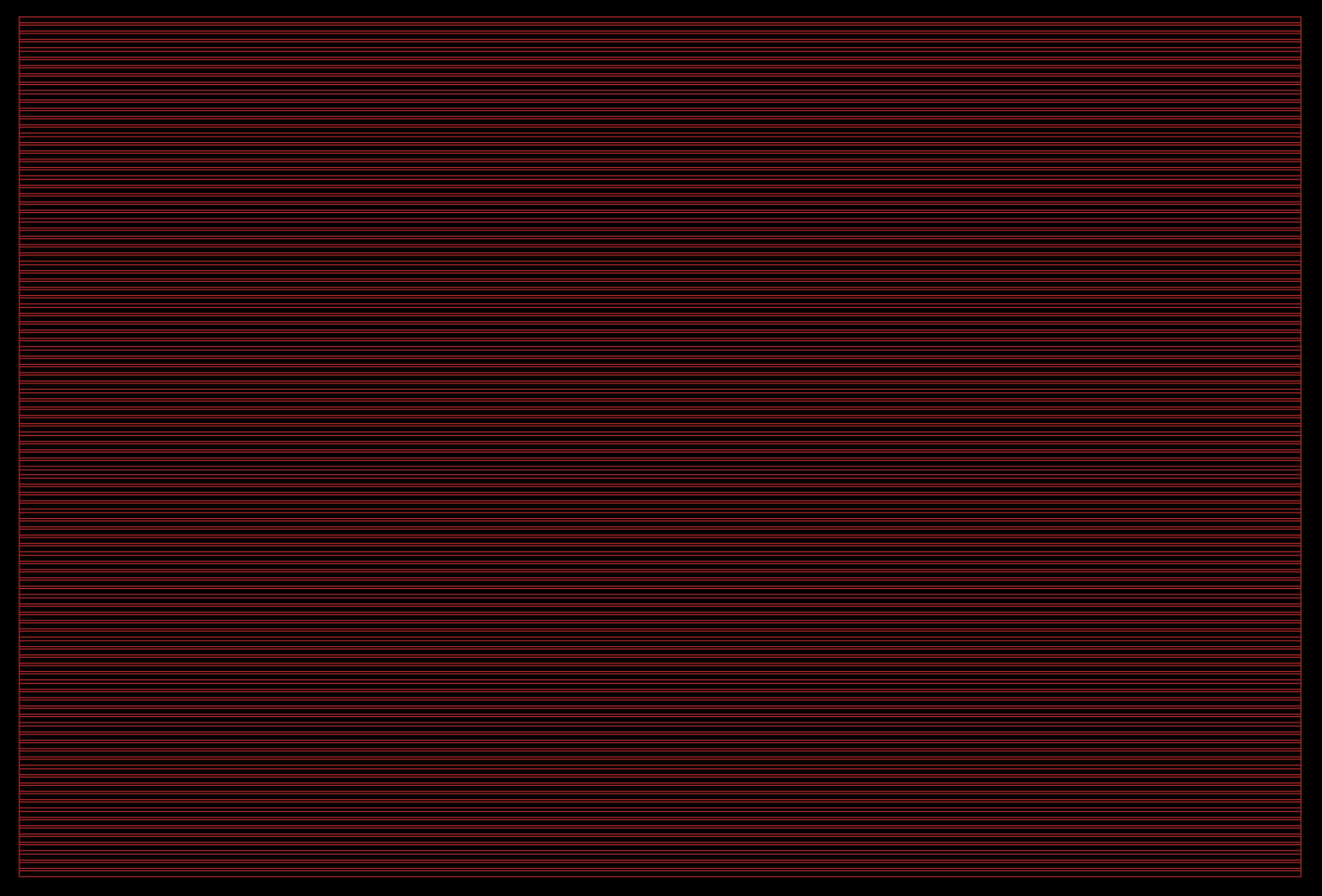}
\caption{Voxelization of 2-m-thick lunar surface in GEANT4 simulations with 100 2-cm-thick cells.}
\label{Voxelization}
\end{center}
\end{figure}

On the first basis, the population and energy spectrum of secondary neutrons induced by the interaction of cosmic ray protons with the lunar soil directly depend on structural parameters such as regolith composition and resulting density. In this sense, a material composition that consists of 43.7 wt.$\%$ oxygen, 0.3 wt.$\%$ sodium, 5.6 wt.$\%$ magnesium, 9 wt.$\%$ aluminum, 21.1 wt.$\%$ silicon, 8.5 wt.$\%$ calcium, 1.5 wt.$\%$ titanium, 0.1 wt.$\%$ manganese, and 10.2 wt.$\%$ iron is initially used in accordance with another study\cite{zaman2022modeling}. Furthermore, a single bulky layer of 3$\times$2$\times$3 m$^{3}$ is taken into consideration to mimic a 2-m-thick lunar surface as described in Fig.~\ref{Lunar_surface_model}.
\label{Simulation_setup}
As illustrated in Fig.~\ref{Lunar_surface_model}, a 2-m-thick single volume of 3$\times$2$\times$3 m$^{3}$ is defined by using a density of either 1.6~\cite{naito2023global} or 1.93~\cite{li2017simulation} g/cm$^{3}$, and the same material composition is utilized for these two different densities. When 0.1 wt.$\%$ hydrogen is introduced into the lunar soil, the mass percentile of the existing oxygen is deducted by the same amount. In all the current simulations, the surrounding medium is vacuum. A pseudo surface detector of 3$\times$0.002$\times$3 m$^{3}$ is set atop the lunar surface so that the albedo neutrons are collected as seen in Fig.~\ref{Lunar_surface_model}. The current lunar surface is voxelized by attempting to compute the generation depth and the initial energy spectrum of the secondary neutrons as shown in Fig.~\ref{Voxelization} where the entire 2-m-thick lunar surface is divided into 100 2-cm cells of 3$\times$0.02$\times$3 m$^{3}$.
Regarding the simulation features in GEANT4, a planar vertical proton beam of 20$\times$20 cm$^{2}$ based on the PAMELA spectrometer that is directed to the center of the lunar surface of 3$\times$2$\times$3 m$^{3}$ is injected along the y-axis as indicated in Fig.~\ref{Lunar_surface_model}. In every simulation, the number of incident protons is 10$^{5}$. All the materials in the current study are defined in accordance with the GEANT4/NIST material database, and the reference physics list used in these simulations is FTFP$\_$BERT$\_$HP. The secondary neutron tracking inside the voxelized lunar surface or at the surface detector is maintained by G4Step, and the tracking information about the secondary neutrons is post-processed via Python. 
\section{Simulation outcomes}
\label{Simulation_outcomes}
\begin{figure}[H]
\begin{center}
\includegraphics[width=8.4cm]{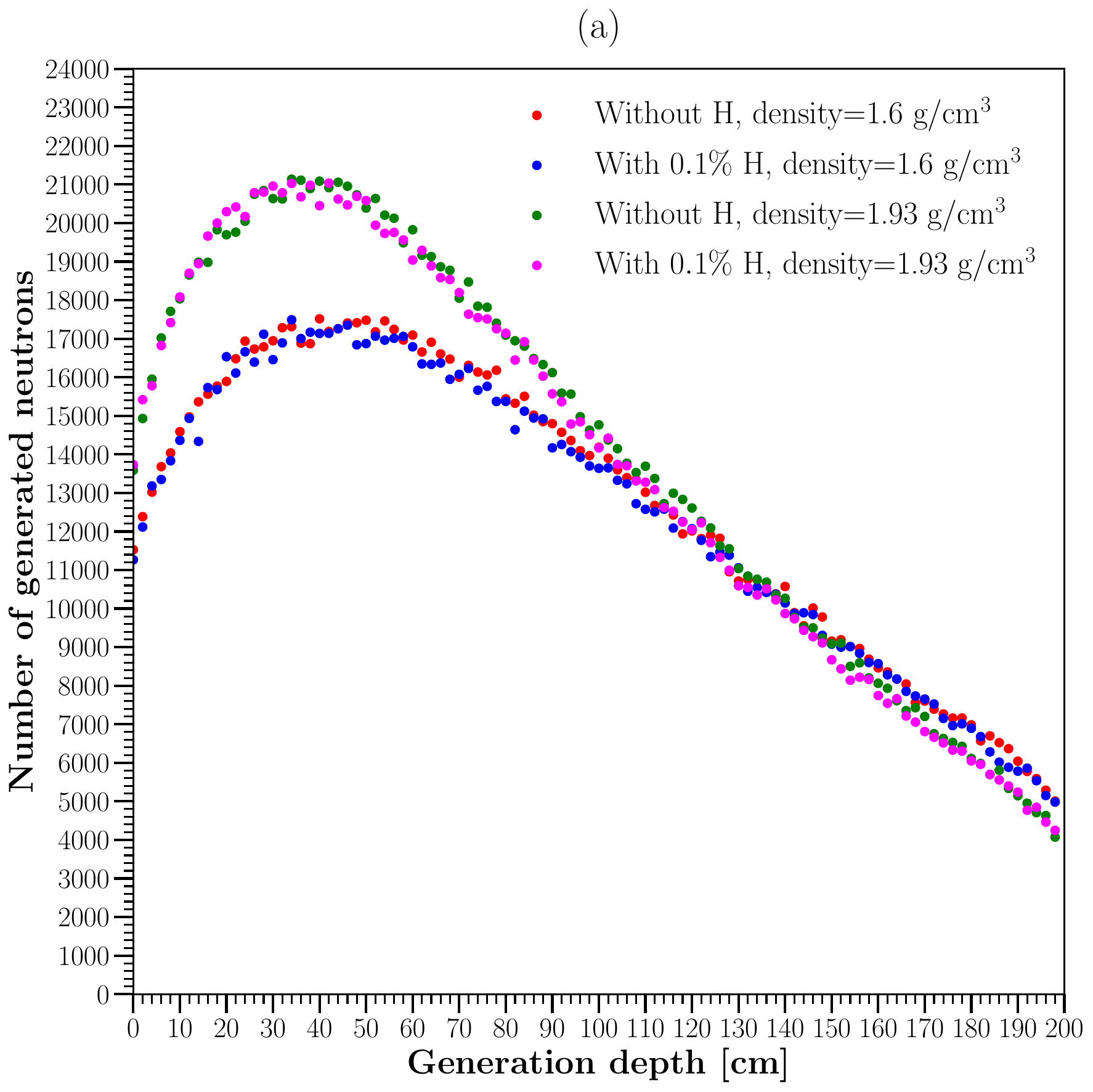}
\includegraphics[width=8.25cm]{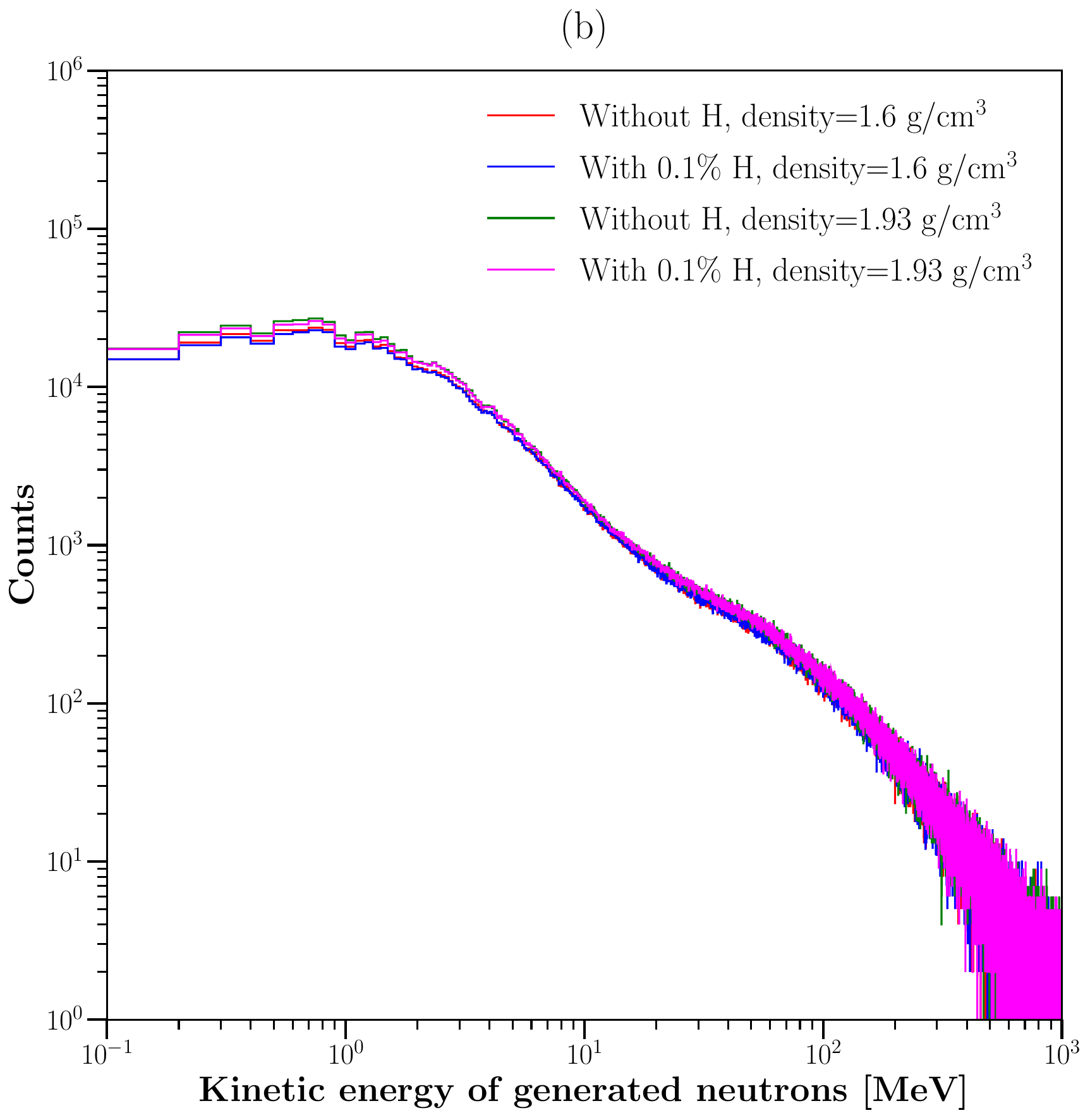}
\caption{Contrast between lunar surfaces of 1.6 and 1.93 g/cm$^{3}$ without/with 0.1 wt.$\%$ hydrogen (a) generation depth of secondary neutrons and (b) initial energy spectrum of secondary neutrons .}
\label{Gen_depth_ener_spec_bulk}
\end{center}
\end{figure}

The profile of the generation depth and the initial energy spectrum for secondary neutrons produced from the interaction of cosmic ray protons with the lunar soil are first obtained by reminding two different densities for this scenario: 1.6 and 1.93 g/cm$^{3}$. Fig.~\ref{Gen_depth_ener_spec_bulk}(a) presents the variation of the secondary neutron population with respect to generation depth in the absence and presence of 0.1 wt.$\%$ hydrogen. As shown in Fig.~\ref{Gen_depth_ener_spec_bulk}(a), introducing 0.1 wt.$\%$ hydrogen into the lunar soil has minimal impact on the generation depth of secondary neutrons. On the other hand, when the surface density is increased from 1.6 to 1.9  g/cm$^{3}$, the peak value increases, and the distribution narrows in terms of generation depth. For both 1.6 and 1.9  g/cm$^{3}$, the maximum generation of secondary neutrons occurs at a depth of approximately 34 cm.
Concerning the initial energy spectrum of secondary neutrons within the single-layer lunar surface, Fig.~\ref{Gen_depth_ener_spec_bulk}(b) depicts the initial energy spectra of secondary neutrons for an energy interval lying between 0.1 and 1000 MeV, and it is noticed that all generated neutrons are considered fast since their initial kinetic energies are all above 0.01 MeV. As observed from Fig.~\ref{Gen_depth_ener_spec_bulk}(b), the initial energy spectra of secondary neutrons have a substantially decreasing trend starting from 1 MeV, and the counts of the cases where the density is 1.93 g/cm$^{3}$ are slightly higher than those of 1.6 g/cm$^{3}$.

\begin{figure}[H]
\begin{center}
\includegraphics[width=8.25cm]{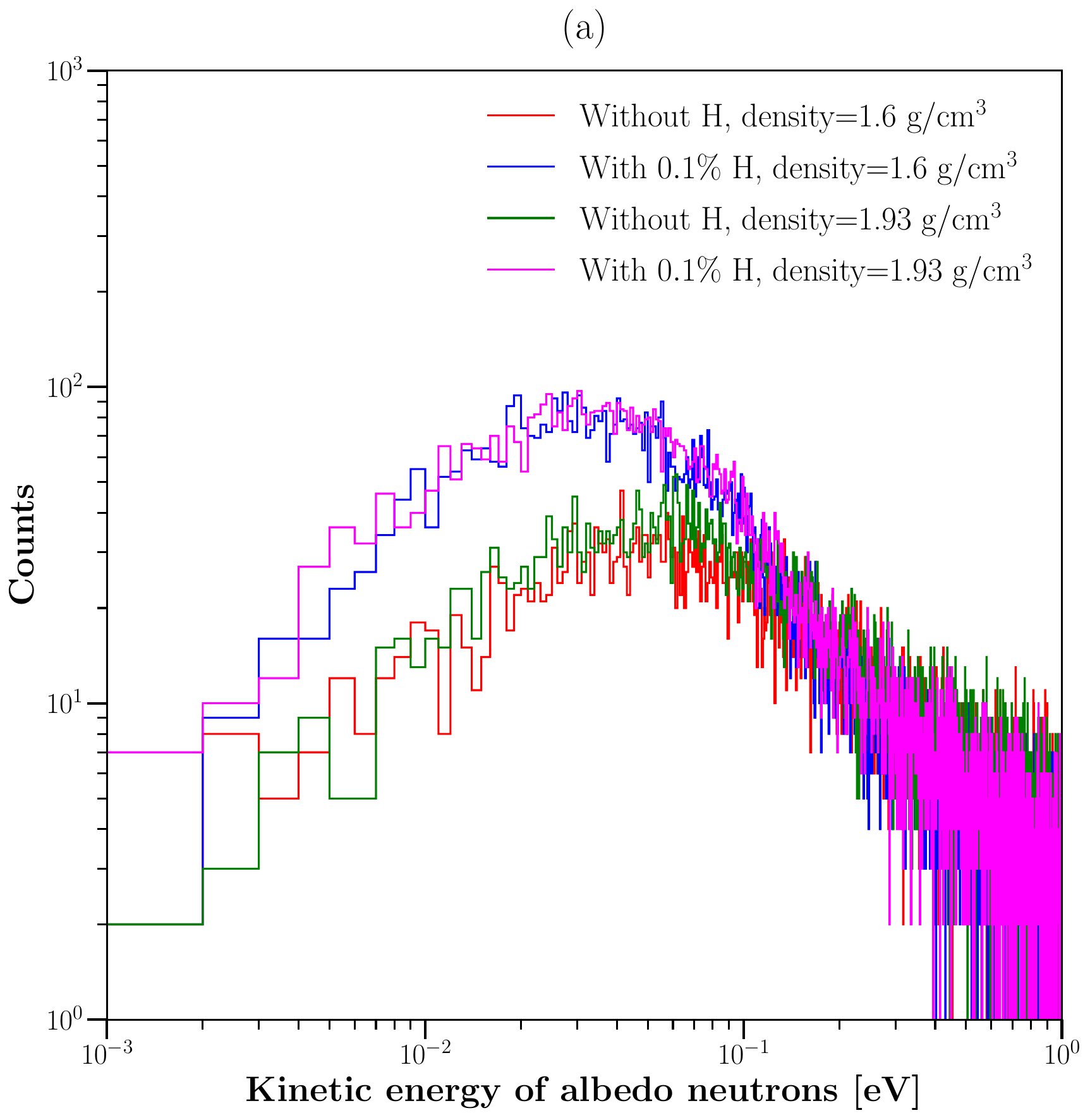}
\includegraphics[width=8.25cm]{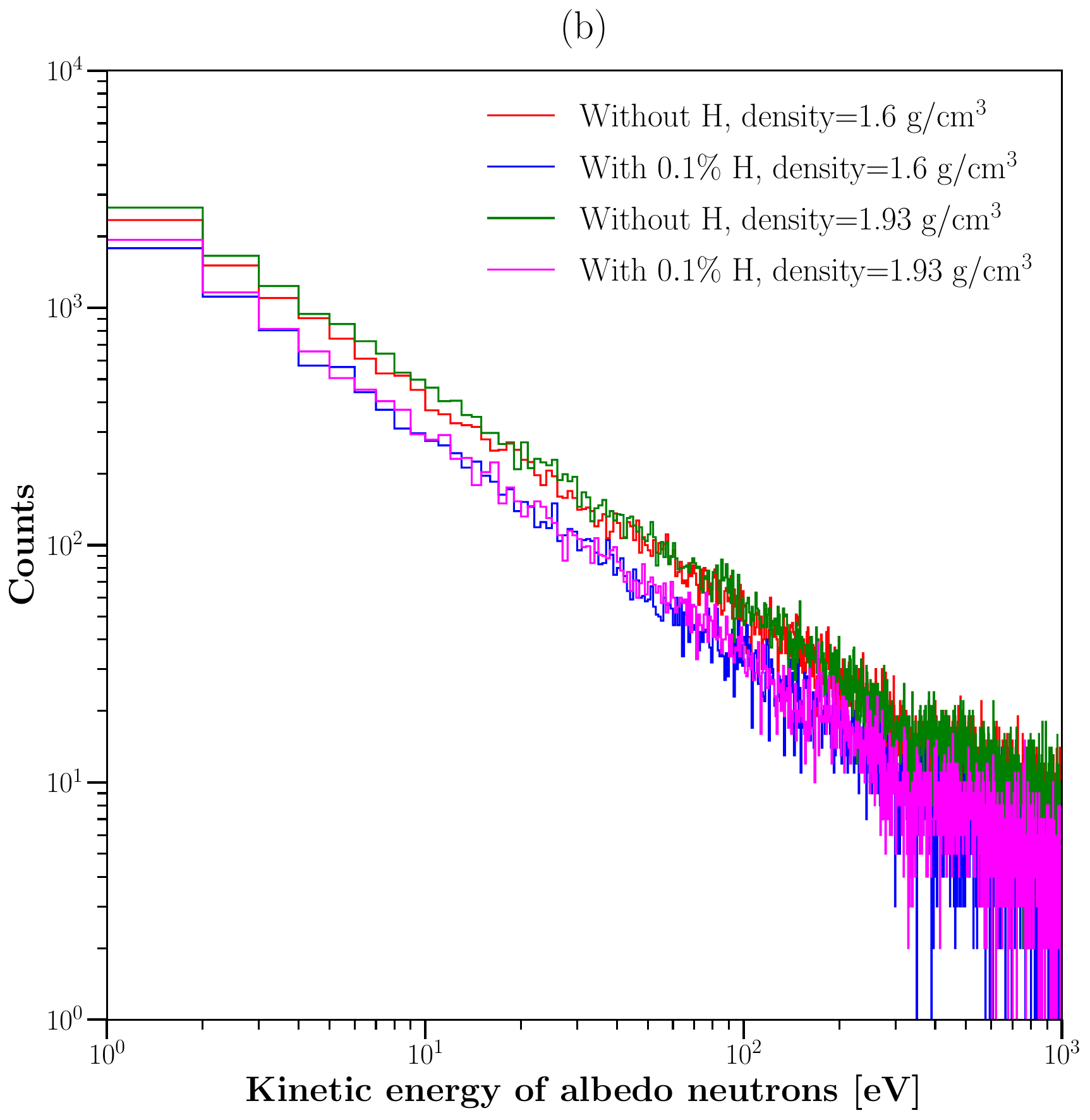}
\includegraphics[width=8.25cm]{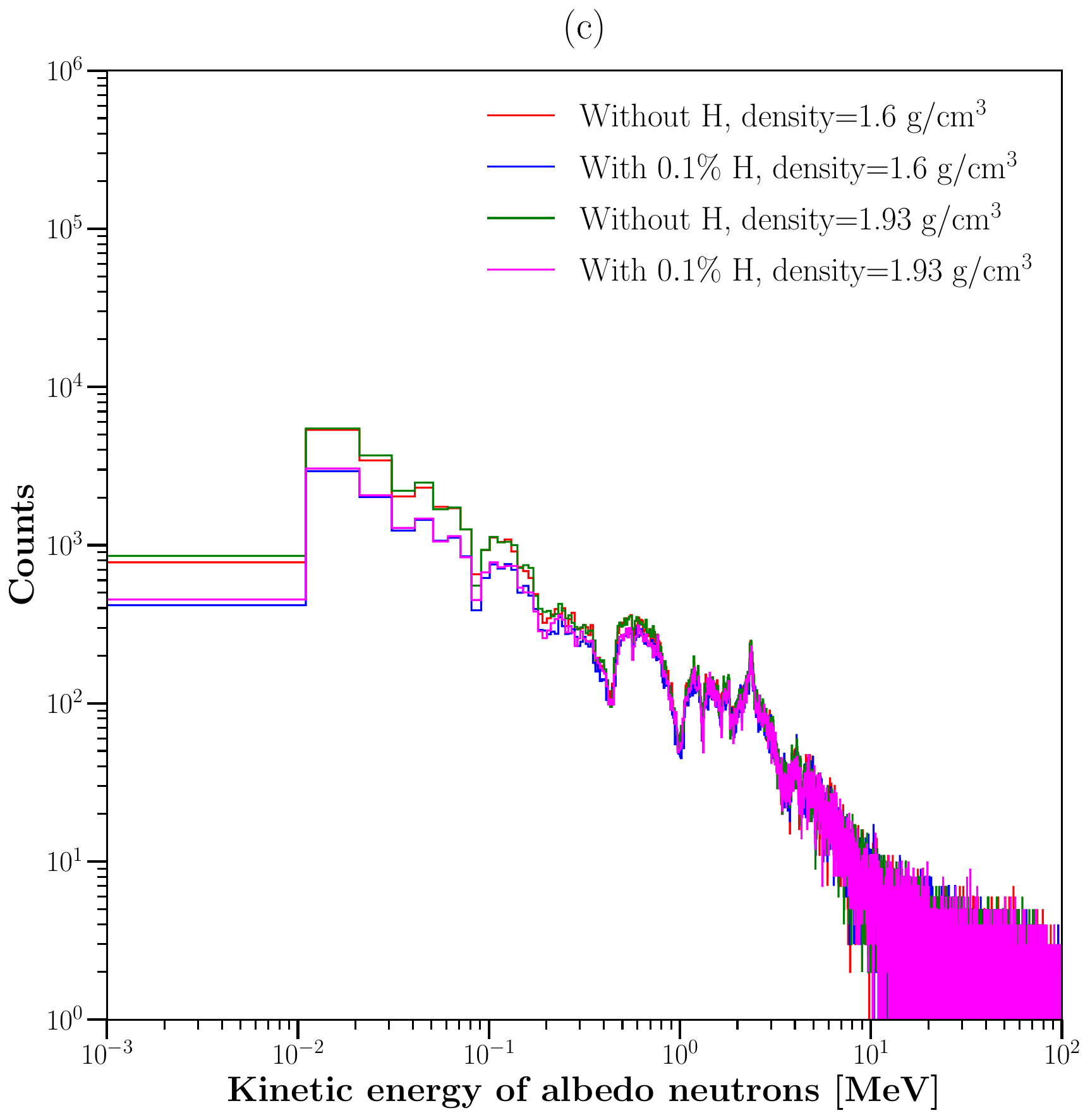}
\caption{Energy spectrum of albedo neutrons without/with 0.1 wt.$\%$ hydrogen atop lunar surfaces of 1.6 and 1.93 g/cm$^{3}$ (a) thermal albedo neutron spectrum, (b) epithermal albedo neutron spectrum, and (c) fast albedo neuron spectrum.}
\label{Albedo_spectrum_bulky}
\end{center}
\end{figure}

In the next step, the albedo neutrons that escape from the defined lunar surfaces are gathered by utilizing a pseudo surface detector. The overall energy spectrum of the albedo neutrons acquired at the surface detector is divided into three energy ranges: thermal ($E\le1$ eV), epithermal (1 eV $< E\le1$ keV), and fast ($E > 1$ keV). The thermal energy spectra of the albedo neutrons for the lunar surfaces of 1.6 and 1.93 g/cm$^{3}$ without and with 0.1 wt.$\%$ hydrogen are exhibited in Fig.~\ref{Albedo_spectrum_bulky}(a). 
Although the presence of hydrogen is hardly observable in the generation depth of the secondary neutrons as depicted in Fig.~\ref{Gen_depth_ener_spec_bulk}(a), it is revealed from Fig.~\ref{Albedo_spectrum_bulky}(a) that the introduction of 0.1 wt.$\%$ hydrogen is traceable above the lunar surface by comparing the thermal energy spectra of the scenarios with and without hydrogen, and the secondary neutrons undergo a non-negligible thermalization process before leaking out from the lunar surface when 0.1 wt.$\%$ hydrogen exists in the lunar soil. Up to 0.1 eV as shown in Fig.~\ref{Albedo_spectrum_bulky}(a), it is seen that the number of the thermal albedo neutrons is sufficiently higher when 0.1 wt.$\%$ hydrogen is present in the lunar soil. Additionally, it is demonstrated that both lunar surfaces of 1.6 and 1.93 g/cm$^{3}$ without hydrogen show more or less similar outcomes, whereas the same resemblance is obtained when the lunar soils with different densities contain 0.1 wt.$\%$ hydrogen. Thus, it is concluded that the introduction of 0.1 wt.$\%$ hydrogen is more influential in the course of the thermalization process compared to the surface density according to the present GEANT4 simulations. Secondly, the epithermal regime in the absence and presence of 0.1 wt.$\%$ hydrogen is checked, the results of which are illustrated in Fig.~\ref{Albedo_spectrum_bulky}(b). Contrary to the previous regime, the cases without hydrogen have more epithermal albedo neutrons within the entire epithermal energy range in comparison with the lunar soils that include hydrogen. Again, it is noted that both lunar surfaces of 1.6 and 1.93 g/cm$^{3}$ without hydrogen exhibit close results, whereas the presence of hydrogen in the lunar soils of different densities results in similar outcomes. In opposition to the thermal regime that shows a parabolic trend along the thermal energy interval in Fig.~\ref{Albedo_spectrum_bulky}(a), the epithermal energy range leads to a linearly decreasing trend in the log-log scale as demonstrated in Fig.~\ref{Albedo_spectrum_bulky}(b). Finally, Fig.~\ref{Albedo_spectrum_bulky}(c) shows the energy spectrum of the fast albedo neutrons for the lunar surfaces of a single volume with different densities in the absence and presence of 0.1 wt.$\%$ hydrogen. In addition to its remarkably different spectral shape, it is observed that the cases without hydrogen have more fast albedo neutrons especially up to 1 MeV; however, all the present cases begin to converge after 1 MeV.

\begin{figure}[H]
\begin{center}
\includegraphics[width=8cm]{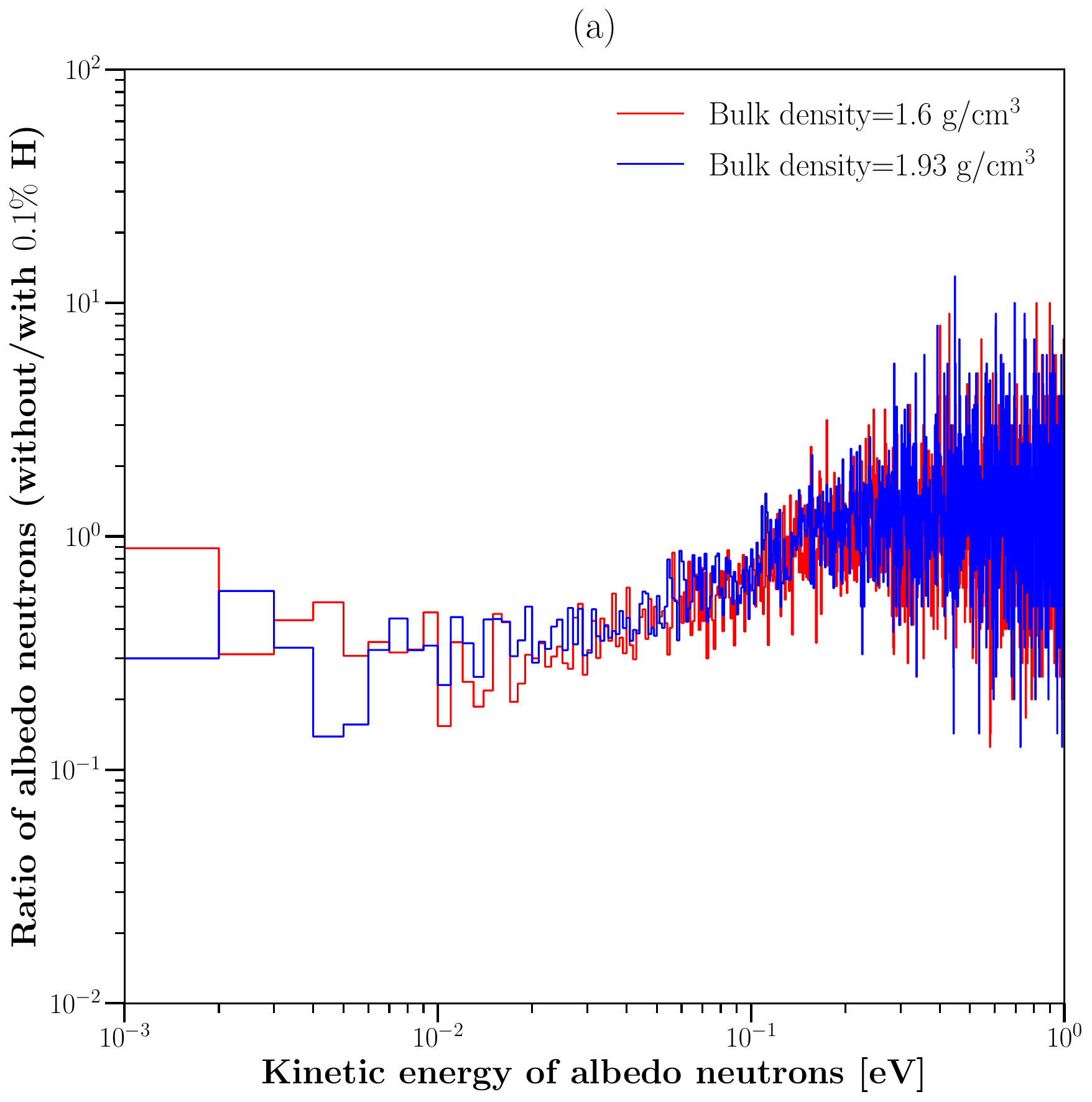}
\includegraphics[width=8cm]{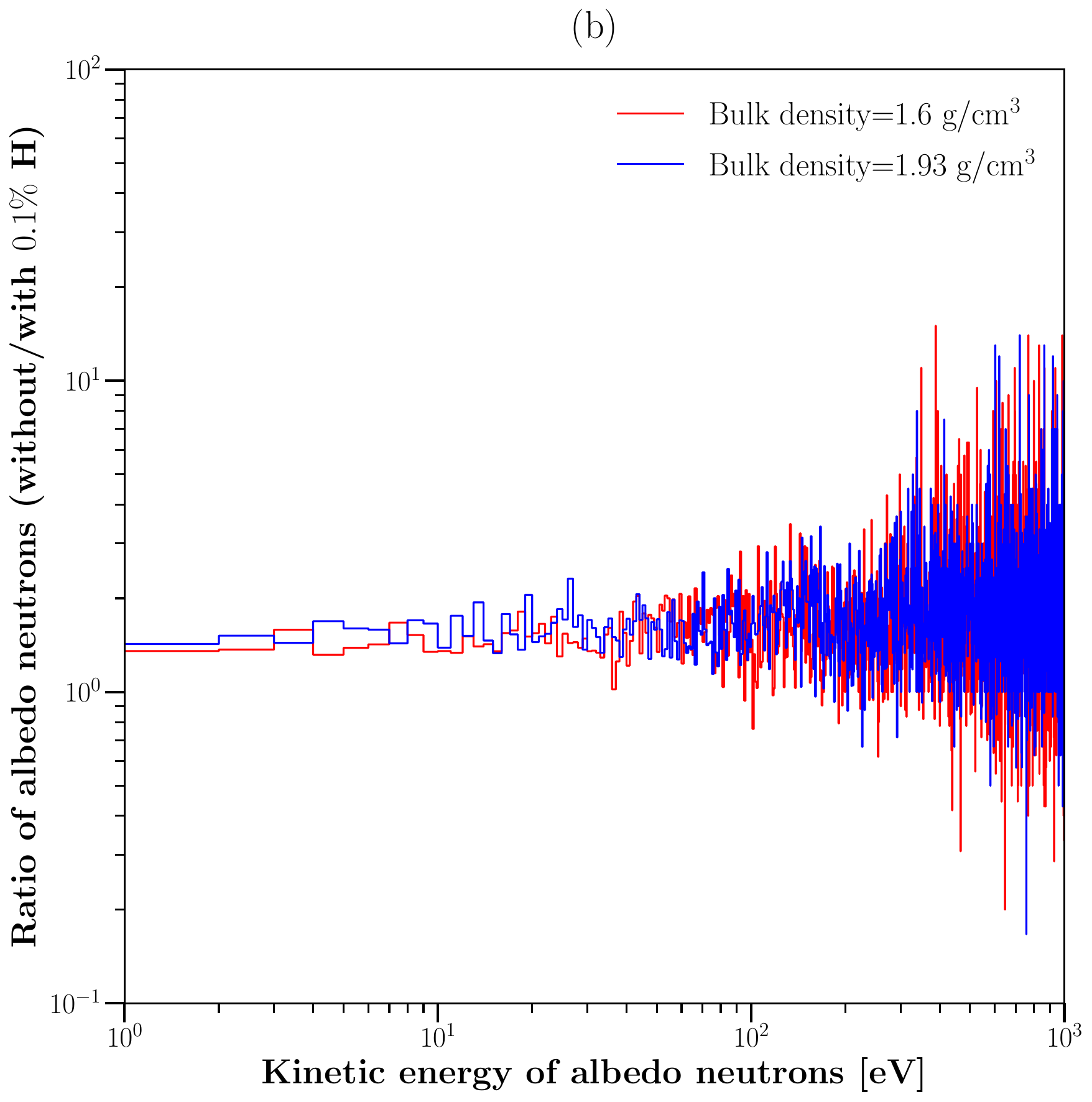}
\includegraphics[width=8cm]{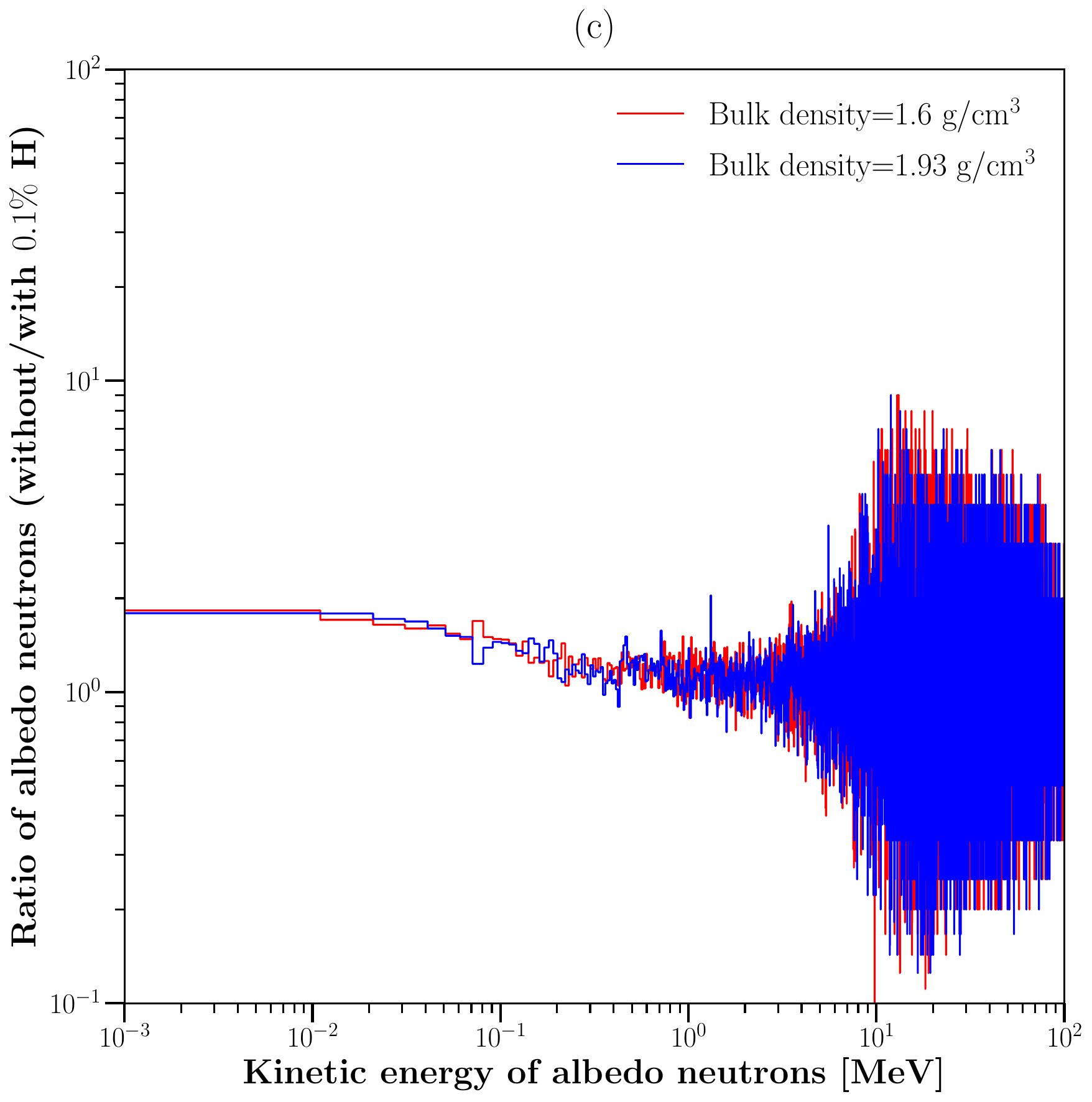}
\caption{Ratio of albedo neutrons without/with 0.1 wt.$\%$ hydrogen atop lunar surfaces of 1.6 and 1.93 g/cm$^{3}$ (a) thermal albedo neutron spectrum, (b) epithermal albedo neutron spectrum, and (c) fast albedo neuron spectrum.}
\label{Ratio_bulk}
\end{center}
\end{figure}

In order to highlight the impact of 0.1 wt.$\%$ on the energy groups of the albedo neutrons at the lunar surface, the ratio between the counts without and with 0.1 wt.$\%$ hydrogen is computed in each specific regime as displayed in Figs.~\ref{Ratio_bulk}(a)-(c). For the thermal regime in the case of a single volume, it revealed that the ratio of the albedo neutrons in the absence and presence of 0.1 wt.$\%$ hydrogen is less than 1 up to 0.1 eV, but it starts to augment after 0.1 eV and converges to 1 when the kinetic energy of the thermal albedo neutrons approaches 1 eV as described in Fig.~\ref{Ratio_bulk}(a). Regarding the epithermal energy range of the albedo neutrons under the same geometrical assumptions, it is noted that the ratio between the populations of the epithermal albedo neutrons without and with 0.1 wt.$\%$ hydrogen is higher than 1 in contrast with the thermal range as indicated in Fig.~\ref{Ratio_bulk}(b). Concerning the fast energy group of the albedo neutrons by using the same geometrical consideration, it is shown that the ratio of the fast albedo neutrons without/with 0.1 wt.$\%$ hydrogen is greater than 1 up to 1 MeV; however, it commences to fall off after 1 MeV and approaches 1 as demonstrated in Fig.~\ref{Ratio_bulk}(c). Additionally, except a very modest deviation, any strong density effect on the ratios of albedo neutrons without and with 0.1 wt.$\%$ hydrogen is not observed at the surface detector under the condition of a single layer as can be seen in Figs.~\ref{Ratio_bulk}(a)-(c).
\section{Conclusion}
In this study, the impact of introducing 0.1 wt.$\%$ hydrogen on the albedo neutron spectra at the lunar surface is investigated via the GEANT4 simulations. By employing a single-layer geometry with two different densities in the absence and presence of 0.1 wt.$\%$ hydrogen, it is observed that a significant thermalization/moderation of secondary neutrons occurs if hydrogen is present in the lunar regolith. The present findings reveal that the existence of 0.1 wt.$\%$ hydrogen markedly increases the number of the thermal albedo neutrons with a noticeable effect up to 0.1 eV by indicating the enhanced neutron thermalization due to the presence of hydrogen. In the epithermal range, the scenarios without hydrogen consistently show higher counts of albedo neutrons across the entire energy spectrum compared to those with hydrogen. This pattern is partially observed in the fast neutron regime where the cases without hydrogen exhibited more fast albedo neutrons up to 1 MeV. It is demonstrated that the introduction of even a small amount of hydrogen substantially alters the albedo neutron spectra by suggesting that neutron spectroscopy is a vital tool for detecting hydrogen and possibly water ice on the Moon. Based on the present GEANT4 simulations, it is suggested that these insights might improve our understanding of lunar surface composition and enhance the strategies for future lunar exploration and resource utilization.
\label{Conclusion}
\bibliographystyle{elsarticle-num}
\bibliography{Moon_surface_hydrogen.bib} 
\end{document}